# $Fe_3O_4$ nano-octahedra and $SnO_2$ nanorods modifying low-Pd amount electrocatalysts for alkaline direct ethanol fuel cells

*Tuani C. Gentil[a,b], Lanna E.B. Lucchetti[a], João Paulo C. Moura[a], Júlio César M. Silva[c], Maria Minichova[b,d], Valentín Briega-Martos[b], Aline B. Trench[a], Bruno L. Batista[a], Serhiy Cherevko[b], Mauro C. Santos[a\*]*

[a] *Centro de Ciências Naturais e Humanas (CCNH), Universidade Federal do ABC (UFABC). Rua Santa Adélia 166, Bairro Bangu, 09210-170, Santo André - SP, Brazil*

[b] *Forschungszentrum Jülich, Helmholtz-Institute Erlangen-Nürnberg for Renewable Energy (IET-2), Cauerstr. 1, 91058, Erlangen, Germany*

[c] *Universidade Federal Fluminense (UFF), Instituto de Química, Grupo de Eletroquímica e Materiais Nanoestruturados, Campus Valonguinho, CEP 24020-141, Niterói, Rio de Janeiro, Brazil*

[d] *Department of Chemical and Biological Engineering, Friedrich-Alexander-Universität Erlangen-Nürnberg, Cauerstr. 1, 91058 Erlangen, Germany*

*Corresponding Author:*

*\*E-mail: mauro.santos@ufabc.edu.br*

# ABSTRACT

This work describes the ethanol oxidation reaction (EOR) in alkaline medium using low-palladium nanoparticle electrocatalysts modified by $Fe_3O_4$ nano-octahedra and $SnO_2$ nanorods. Operation studies on an alkaline direct ethanol fuel cell (ADEFC) were conducted using the developed electrocatalysts, and stability studies were performed using the advanced scanning flow cell (SFC) technique coupled to inductively coupled plasma mass spectrometry (online SFC-ICP-MS). The EOR was catalyzed by single (Pd/C and commercial Pd/C Alfa Aesar) and by synthesized binary/ternary electrocatalysts, in which $Fe_3O_4$ and $SnO_2$ nanostructures partially replaced the high-cost noble metal. The $PdFe_3O_4$/C was identified as the most promising synthesized material in the electrochemical studies, exhibiting the highest mass activity (1426 mA $mg^{-1}$ Pd) by cyclic voltammetry (CV), followed by the binary $PdSnO_2$/C (1135 mA $mg^{-1}$ Pd), and by the ternary (1074 mA $mg^{-1}$ Pd). This enhancement was attributed to the bifunctional mechanism enabled by $Fe_3O_4$ and $SnO_2$, therefore reducing poisoning and improving the EOR. Moreover, the operating results revealed that $PdFe_3O_4$/C showed the highest power density among the synthesized materials (31 mW $cm^{-2}$ at 70 °C), even with a ~45% reduction in Pd content compared to the commercial. XPS results showed that the Pd $3d_{5/2}$ and $3d_{3/2}$ peaks for $PdFe_3O_4$/C, $PdSnO_2$/C, and $PdFe_3O_4SnO_2$/C were shifted by ~0.5 eV to higher binding energies compared to Pd/C, indicating a loss of electron density in Pd due to strong metal–oxide interactions. These interactions led to a downward shift in the d-band center of Pd, weakening the Pd-adsorbed bonds, facilitating the desorption of intermediates, and improving the catalyst tolerance to toxic species. Furthermore, the higher proportion of Pd oxides in the binary and ternary materials appeared to contribute to the supply of oxygenated species required for the oxidation of EOR intermediates. Thus, the observed enhancement resulted from synergistic bifunctional and electronic effects. Finally, online SFC-ICP-MS studies showed that $Fe_3O_4$ nano-octahedra contribute to the enhanced stability of the electrocatalyst, as $PdFe_3O_4$/C exhibits reduced Pd dissolution and no Fe dissolution.

# INTRODUCTION

Ethanol is one of the main biofuels studied in electrocatalysis due to its renewable characteristics and low toxicity [1]. The volumetric energy density of ethanol is approximately 23 MJ dm$^{-3}$, higher than methanol (~15 MJ dm$^{-3}$) and hydrogen (~5 MJ dm$^{-3}$). When applied in fuel cells, its theoretical efficiency is 97%, more significant than that of those operating with $H_2$ (83% ) [1,2].

Beyond its lower volumetric energy density relative to ethanol, hydrogen presents additional challenges for $H_2$–$O_2$ fuel cells, particularly due to its intrinsically explosive nature. In the Brazilian context, ethanol represents a particularly attractive alternative [1,3], as the country is the world's largest producer of sugarcane-based ethanol—a sector that has experienced substantial expansion over the past three decades. Given this background, ethanol stands out as a promising sustainable energy source [1,3–6].

Assuming the transfer of 12 electrons for each oxidized ethanol molecule, obtaining 8.1 kW kg$^{-1}$ of energy density is possible. However, the challenge for ethanol oxidation to $CO_2$ remains in cleaving the C-C bond [7,8]. According to mechanisms discussed in the literature, the main products of incomplete ethanol oxidation are acetic acid and acetaldehyde [9]. Another known mechanism can lead to carbon monoxide (CO) formation, responsible for catalytic poisoning due to its strong interaction with the electrocatalyst's metallic active sites, preventing new ethanol molecules from being adsorbed [9,10].

Nanostructured materials based on palladium [11–14] and platinum [13,15] are usually active for ethanol electrooxidation and commonly used as electrocatalysts for application in direct ethanol fuel cells (DEFC). In alkaline direct ethanol fuel cells (ADEFC), Pd has advantages over Pt, reaching up to four times more activity in ethanol oxidation reaction (EOR) [16]. In addition, promising results have been presented in the literature regarding current and power density catalyzed by Pd nanoparticles in alkaline direct liquid fuel cells (ADLFC) [9,17,18]. However, Pd and Pt, although promising, are costly noble metals [6,19]. In 2023, the cost of Pd surpassed that of Pt, reaching $1315.83/oz versus $968.60/oz for Pt [20].

Thus, to reduce the total amount of costly precious metals, Pd-based materials combined with auxiliary metals (metallic co-catalysts and metallic oxides) have been studied in the last decades [21–26]. The co-catalysts contribute to cost reduction and improve stability and tolerance to poisoning, enhancing the practicality of fuel cell applications. One of the discussions in the literature [25,27,28] is the supply of oxygenated species, which can facilitate the oxidation of intermediate compounds according to the bifunctional mechanism, in which the metallic base acts as a nucleus for the adsorption and dehydrogenation of organic species. Meanwhile, the auxiliary metal, activated by water, allows the formation of oxygenated species at a lower potential than the base metal. When these oxygenated species are formed near CO adsorption sites, the oxidation of CO to $CO_2$ is facilitated, and catalytic poisoning is reduced.

The formation of alloys between base and secondary metals can enhance electrocatalytic activity due to the electronic effects that modify the energy levels of the Pd *d*-band. This leads to weaker CO adsorption, facilitating its subsequent oxidation to $CO_2$ [28].

Studies using $Pd_xSn_y$/C showed that the Pd-Sn alloy can favor EOR in alkaline medium [4]. Sn can favor the oxidation of intermediate species adsorbed on the Pd surface, releasing their active sites to oxidize a new ethanol molecule. Studies involving binary electrocatalysts based on Pd with secondary metals such as Au, Ir, and Ru revealed tolerance to ethanol impurities and EOR intermediate products, resulting in materials with more excellent stability and lower onset potential [29–33].

Using oxides with noble metals favors the bifunctional mechanism in acidic and alkaline media, enhancing electrocatalytic activity. Gentil *et al.* [6] studied the addition of ceria oxides ($CeO_2$ nanorods) combined with PtSn for EOR in an acid medium, showing that materials with a higher load of $CeO_2$ nanorods favored CO oxidation, according to CO-stripping experiments. Pinheiro *et al.* [34] conducted CO-stripping studies in an alkaline medium using Pt-based materials combined with $CeO_2$ nanorods. The results showed that catalysts containing $CeO_2$ nanorods exhibited higher electrochemically active surface area (ECSA) values compared to those obtained with commercial Pt/C. Iron oxide is also an excellent co-catalyst for EOR, with recent studies using $Fe_3O_4$ in combination with Pd/C electrocatalysts revealing that EOR was favored in alkaline medium [35,36]. Hongmin Mao *et al.* [37] explored using $SnO_2$ nanoparticles with PdSn for EOR in an alkaline medium, and the authors suggested that the OH adsorption is favored on the $SnO_2$ surface, promoting the oxidative removal of $CO_{ads}$ in Pd. Considering the electrocatalysis-favoring effects of nanorod morphology and $SnO_2$ composition, in this work, we combine these two characteristics, synthesizing $SnO_2$ nanorods, to evaluate the EOR contribution of Pd-based materials.

Therefore, this work proposes an investigation of EOR in an alkaline medium catalyzed by materials based on low palladium amount nanoparticles supported on carbon with Fe and Sn metallic oxides as co-catalysts. The synthesized materials were also studied as anodic catalysts under ADEFC operating conditions. The morphologies of $Fe_3O_4$ nano-octahedra and $SnO_2$ nanorods were explored, supported by the promising effects of morphology and composition found in the literature, but have yet to be studied according to our knowledge, under the conditions proposed in this work. The stability of the most promising material for ethanol oxidation was further investigated using a flow scanning cell (SFC) coupled with inductively coupled plasma mass spectrometry (ICP-MS ).

**EXPERIMENTAL SECTION**

**Synthesis of electrocatalysts**

Palladium nanoparticles were synthesized using sodium borohydride as a reducing agent [6,34,38–40], and Vulcan XC-72 Carbon was used as support. Palladium(II) nitrate dihydrate ($Pd(NO_3)_2·2H_2O$, Sigma-Aldrich) was used as the metal precursor and dissolved in a 1:1 (v/v) mixture of water and isopropanol. To prevent the agglomeration of metallic nanoparticles, cetrimonium bromide (CTAB) was employed at a molar ratio of 1:20 (CTAB:metal). A sodium borohydride solution ($NaBH_4$, Sigma-Aldrich) was freshly prepared, with a concentration exceeding 10 mol L$^{-1}$, in 0.01 mol L$^{-1}$ potassium hydroxide (KOH, Synth). It is important to prepare $NaBH_4$ in an alkaline media, as it degrades readily under neutral or acidic conditions. Following the addition of the reducing agent, the mixture was stirred for 30 minutes. Subsequently, Vulcan XC-72 carbon (Cabot) was introduced as the support material, and the resulting suspension was mechanically stirred for an additional 150 minutes. The final product was washed with a 1:1 (v/v) mixture of ethanol and ultrapure water, then dried in an oven at 90 °C. The nominal mass ratio of the Pd/C simple material was 20% by mass of Pd and 80% by mass of carbon support.

To prepare the binary electrocatalysts, 45% of the Pd mass was replaced by $Fe_3O_4$, resulting in the $PdFe_3O_4$/C material, and by $SnO_2$, resulting in the $PdSnO_2$/C material. The nominal mass ratio of the binary materials was Pd: oxide (3:2). On the other hand, for the ternary preparation, 45% of the Pd mass was replaced by $Fe_3O_4$ and $SnO_2$, resulting in the $PdFe_3O_4SnO_2$/C material. The nominal mass ratio, in this case, was Pd:$Fe_3O_4$:$SnO_2$ (3:1:1). Binary and ternary materials were prepared to have 85% carbon support by mass. The $Fe_3O_4$ nano-octahedra and the $SnO_2$ nanorods were previously synthesized and impregnated in the Pd/C material [41].

In the impregnation procedure [34,42], the proper amounts of Pd/C were weighed and then dispersed in 50 mL of ultrapure water with the aid of a magnetic stirrer for 1 h. A dispersion of the proper amounts of oxides in 20 mL of ultrapure water was prepared to synthesize the Pd:oxide (wt%) electrocatalysts. The mixture was subjected to continuous stirring using a magnetic stirrer for 24 hours. For the electrocatalysts containing $Fe_3O_4$, stirring was carried out using a platform shaker due to the magnetic properties of magnetite. After this step, the synthesized electrocatalysts were dried in an oven at 90 °C.

**Synthesis of $Fe_3O_4$ nano-octahedra**

The hydrothermal route was used for the synthesis of $Fe_3O_4$ nano-octahedra, in which 0.328g of $FeSO_4.7H_2O$ was dissolved in 10 mL of ultrapure water, and 22 mL of propylene glycol was added drop by drop using a burette. Mechanical agitation was carried out with a platform shaker for 1 hour. After that, 1.6 g of NaOH dissolved in 5 mL of ultrapure water was added and stirred again for 10 minutes. The content was heated in a muffle inside a Teflon-coated autoclave flask at a speed of 10°/min to 200°C and maintained at this temperature for 20 hours. Cooling was performed at room temperature, and the solid was washed with water: ethanol (1:1), centrifuged, and taken to a vacuum oven at a temperature of 60 °C for drying [35].

**Synthesis of $SnO_2$ nanorods**

The hydrothermal route was used to synthesize $SnO_2$ nanorods, in which 4.5 mmol of $SnCl_2$ was dissolved in 30 mL of ultrapure water in a beaker. Magnetic stirring was carried out until a white paste was obtained. The beaker's contents were transferred to an autoclave flask, and 3.75 mL of $H_2O_2$ (30% v/v) and 27 mmol of NaOH were added. Mechanical agitation was carried out with a platform shaker until the NaOH dissolved. The mixture was placed in a Teflon-lined autoclave and heated in a muffle furnace at a rate of 10 °C/min up to 200 °C, where it was maintained for 30 hours. After natural cooling to room temperature, the resulting solid was successively washed with ultrapure water and ethanol, followed by centrifugation, and then washed again with ethanol. Drying was done in a vacuum oven at 50°C [43].

**Stability Measurements Using Online SFC-ICP-MS**

The Pd dissolution was investigated by mapping stability measurements in the presence of ethanol, using an in-house design of SFC [44] coupled to the ICP-MS (Online ICP-MS - Nexion 300X, PerkinElmer). A glassy carbon (GC) rod (HTW SIGRADUR G) was used as the counter electrode, and a commercial Ag/AgCl (3 M KCl, Metrohm) was used as the reference electrode (calibrated to RHE prior measurements), both attached to the inlet and outlet tubes of SFC, respectively. A GC substrate (SIGRADUR G, HTW 25 cm$^2$) with studied electrocatalysts was used as the working electrode. Aliquots of 0.3 µL of the electrocatalyst inks were dropped on the GC substrate, reaching approximately 20 µg cm$^{-2}$ total metal loading. The spots were then air-dried at room temperature. The radius of each spot was determined using a laser microscope (Keyence VK-X250), ranging between 600 and 800 µm, and all electrochemical and dissolution data were normalized against the particular geometric surface area. A simplified scheme of the basic operating principles of the online SFC-ICP-MS is shown in the supporting information (**Fig. S1**).

**Electrochemical characterization and EOR studies in alkaline medium**

The electrochemical experiments were performed with an Autolab PGSTAT 302N potentiostat/galvanostat. An electrochemical cell with three electrodes was used: glassy carbon as the working electrode, with an area of 0.196 cm$^2$ where the electrocatalyst dispersions were deposited; Pt as the counter electrode with 2 cm$^2$; Saturated Calomel Electrode (SCE) as the reference electrode. In the electrochemical cell, 1.0 mol L$^{-1}$ KOH was used as the electrolyte, and the experiments were carried out at 25 °C in a deoxygenated medium, purging nitrogen in the electrolyte for approximately 10 minutes before each measurement. For the EOR, the same experimental conditions were used, and 1.0 mol L$^{-1}$ of ethanol was applied to the electrochemical cell. For such experiments, a thin layer of 20 µL of electrocatalyst dispersions was applied to the working electrode. To prepare the dispersion, 8 mg of material was dissolved in 1 mL of ultrapure water and isopropanol (80:20, respectively). The dispersion was taken to the ultrasound for 2 minutes at 20% amplitude. After this process, 20 µL of this dispersion was added to the working electrode surface. The surface was dried with a tungsten lamp, and after this process, 20 µL of Nafion® solution in water (1:100 v/v, respectively) was added, and drying was performed again under the same conditions.

Cyclic voltammetry was recorded using a sweep rate of 20 mV s$^{-1}$ in 1.0 mol L$^{-1}$ KOH in the absence and presence of ethanol in a potential window of -1.0 to +0.2 V vs. SCE. Chronoamperometric measurements were also performed at potentials of -0.3V vs. SCE, and linear sweep voltammetry (LSV) experiments using a sweep rate of 0.5 mV s$^{-1}$ were used to assess the Tafel slope.

**Electrochemically Active Surface Area (ECSA)**

The electrocatalysts' ECSA [4,5] was determined by CO-stripping analysis, performed at room temperature using 1.0 mol L$^{-1}$ KOH as electrolyte. Ultra-pure N$_2$ (99.99%) was purged into the solution for 15 minutes, then the modified electrode was immersed in the solution, and CO was bubbled through for 10 min. Excess CO was removed by bubbling N$_2$ for 15 min, and finally, the measurement was made in the potential window from -0.6 V to 0.1 V vs. SCE at a sweep rate of 20 mV s$^{-1}$.

**Experiments in Alkaline Direct Ethanol Fuel Cells (ADEFC)**

For the single fuel cell experiments, membrane electrode assemblies (MEAs) were previously prepared as described in the previous work [5]. The electrolyte used is made up of Nafion® 117 membranes, which undergo previous treatment to remove organic and inorganic contaminants from the membranes. The treatment consists of washing and specific activation for the operating environment. After this process, a geometric area of 5 cm² of the Teflon-coated carbon fabric was deposited with the synthesized anodic electrocatalysts based on Pd nanoparticles and the commercial cathodic electrocatalyst Pt/C (Alfa Aesar). The materials were dried at 70 °C for 2 hours for subsequent preparation of the MEAs, using a mass of 5 mg of electrocatalyst per square centimeter. The MEAs were made by hot pressing in a hydraulic press at a temperature of 120 °C and a pressure of 5 tons for 7 minutes. The MEAs were prepared using electrocatalysts with 5% Nafion solution in 1.0 mg cm$^{-2}$ electrocatalyst loading.

The single fuel cell (Electrocell® - Fuel Cell Energy) used in the experiments was detailed in the previous work [5]. The experiments were carried out with an Autolab PGSTAT 302N potentiostat/galvanostat. The anode was supplied with a 2 mol L$^{-1}$ ethanol solution in 1 mol L$^{-1}$ KOH as the fuel, while the cathode was fed with oxygen gas at a flow rate of 500 mL min$^{-1}$. The studies were carried out at different operating temperatures (40 °C, 50 °C, 60 °C, 70 °C, 80 °C and 90 °C), at a pressure of 2 bar (200 kPa) and a flow of 1.0 mL min$^{-1}$ of fuel at the anode. The oxygen humidifier was set at 85 °C with 200 mL min$^{-1}$ flow at the cathode.

**Physical-chemical characterization**

The electrocatalysts were physically characterized by X-ray diffraction (XRD), inductively coupled plasma mass spectrometer (ICP-MS), High-resolution transmission electron microscopy (HR-TEM), Scanning electron microscopy (SEM), Dispersive energy spectroscopy (EDS), and X-ray photoelectron spectroscopy (XPS). Detailed information regarding the equipment, sample preparation, measurement parameters, and data processing for the physical-chemical characterization is provided in the Supplementary Material.

**RESULTS AND DISCUSSION**

**Physical-Chemical Characterizations**

The chemical composition of the electrocatalysts Pd/C, PdFe$_3$O$_4$/C, PdSnO$_2$/C, and PdFe$_3$O$_4$SnO$_2$/C was determined by EDS (semi-quantitative) and by ICP-MS (quantitative), according to the arithmetic means shown in **Tab. 1**. A nominal mass ratio for the simple Pd/C was 20% Pd and 80% C. The electrocatalysts containing metal oxides were synthesized to reduce the palladium metallic load by replacing it with low-cost oxides, with a nominal mass ratio of 9% Pd, 6% oxides (Note: in the ternary material, the nominal mass ratio between the oxides was 1:1), and 85% C. The palladium and metallic oxide contents for the binary and ternary materials were close to the nominal mass ratio according to the quantitative results of ICP-MS.

**Tab. 1.** Elemental composition estimated by EDS and ICP-MS of electrocatalysts.

| Electrocatalyst | EDS (%) | | | | Oxides | | ICP-MS (%) |
|---|---|---|---|---|---|---|---|
| | C | Pd | Fe | Sn | Fe$_3$O$_4$ | SnO$_2$ | Pd |
| Pd/C | 80.1 ± 2.2 | 19.9 ± 2.2 | (-) | (-) | (-) | (-) | 16.0 ± 0.10 |
| PdFe$_3$O$_4$/C | 89.4 ± 1.5 | 6.4 ± 1.1 | 4.3 ± 1.5 | (-) | 5.9 | (-) | 10.7 ± 0.03 |
| PdSnO$_2$/C | 84.6 ± 1.8 | 7.6 ± 1.4 | (-) | 7.9 ± 0.6 | (-) | 10.0 | 10.9 ± 0.30 |
| PdFe$_3$O$_4$SnO$_2$/C | 89.4 ± 2.1 | 6.0 ± 2.1 | 1.9 ± 0.2 | 2.7 ± 0.3 | 2.6 | 3.5 | 11.0 ± 0.03 |

Elemental mapping was performed using SEM/EDS to characterize the spatial distribution of chemical elements on the surface of the materials. SEM allowed the acquisition of micrographs that detail the morphological and topographic characteristics of the material, while EDS allowed the qualitative and semi-quantitative identification of elements, and together they provided critical insights into the chemical heterogeneity of the sample and compositional uniformity [45], thus contributing to a deeper understanding of the physicochemical properties of the surface and elemental dispersion of C, Pd, Fe, Sn, and O.

As shown in **Fig. 1**, SEM elemental mapping confirms the presence of Pd, Fe, O, and C in the PdFe$_3$O$_4$/C electrocatalyst. The dispersion of Pd and Fe$_3$O$_4$ nano-octahedra corresponds well, indicating the successful synthesis, and also suggests the nano-octahedral morphology of Fe$_3$O$_4$, which was confirmed with other microscopy techniques, as well as the anchoring on the Vulcan XC-72 carbon support. The EDS dispersion results revealed the existence of Pd, Fe, O, and C, confirming the successful preparation and composition of the material. No large agglomerations related to the metals were observed, and they show that Pd and Fe are close together in similar regions, which may increase the electrochemically active potential of the electrocatalyst due to area effects and electronic effects [46]. Fe and O were observed in similar regions, confirming the composition of Fe$_3$O$_4$.

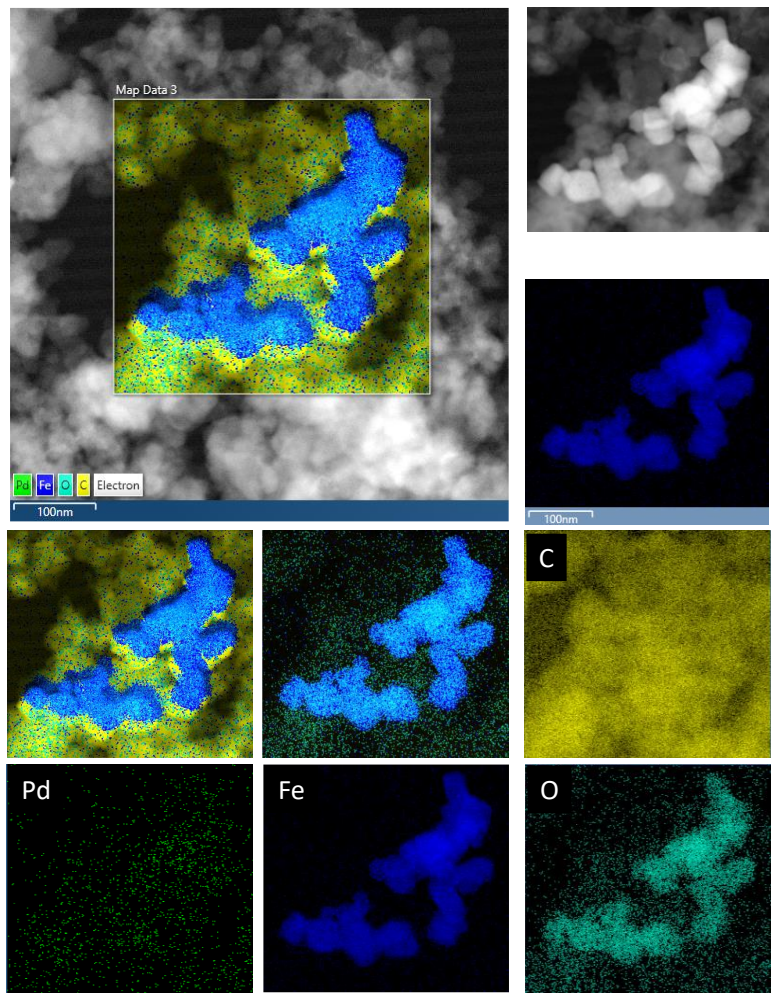

**Fig.1**. Mapping performed by SEM/EDS of the electrocatalyst PdFe$_3$O$_4$/C.

**Fig. S2** shows the surface mapping of the binary PdSnO$_2$/C. The results confirmed the material's elemental composition and the nanostructures' distribution on the carbon support. The surface mapping of the ternary material PdFe$_3$O$_4$SnO$_2$/C is shown in **Fig. 2**, confirming the catalyst composition, and it is possible to identify the Pd, Fe, Sn, O, and C and suggest the characteristic morphologies Fe$_3$O$_4$ (nano-octahedra) and SnO$_2$ (nanorods). It was also noted that the elements are close to each other, which might enhance the catalytic activity [36]. The semi-quantitative result obtained by EDS showed that Pd is in relatively low concentration compared to Fe and Sn, possibly due to lower homogeneity of the material. Nonetheless, a more representative and quantitative analysis by ICP-MS revealed that the electrocatalyst presents a Pd amount close to the starting mass ratio.

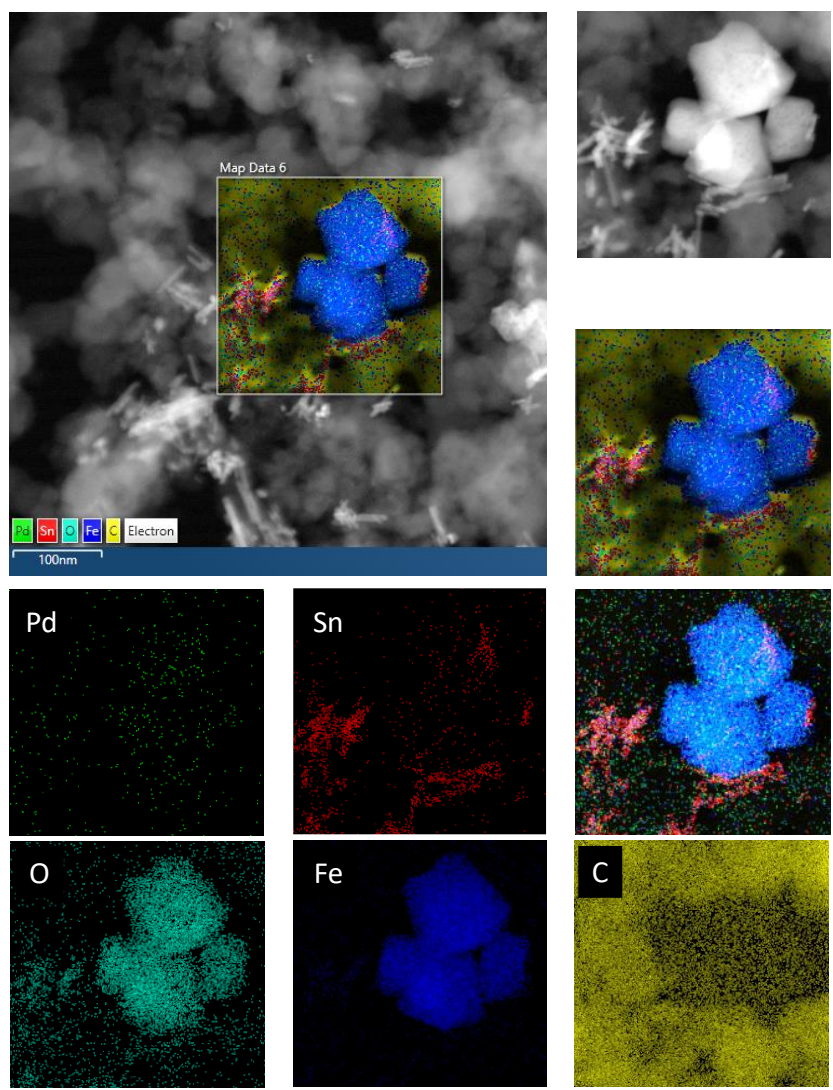

**Fig. 2.** Mapping performed by SEM/EDS of the electrocatalyst PdFe$_3$O$_4$SnO$_2$/C.

The X-ray patterns of Pd/C, PdFe$_3$O$_4$/C, and commercial Pd/C Alpha Aesar (Pd/C AA) materials were carried out in previous studies [5]. They indicated the presence of metallic Pd (PDF 5-681) as confirmed by reflections (111) and (200) at 40° and 46°, respectively, compared to the Pd/C AA, corroborating the literature [16]. The XRD patterns of all electrocatalysts (**Fig. 3a**, **Fig. 3b**, and **Fig. 3c**) are by the Joint Committee on Powder Diffraction Standards (JCPDS) and show the presence of metallic Pd (PDF 5-681) with a cubic structure fcc(ccp)-Cu, confirmed at 40° and 46° [4]. The PDF 19-629 standard was used to identify the diffraction signals that can be correlated with a cubic phase (fcc, Fd-3 m space group) of the magnetite structure and Fe$_3$O$_4$ phases and crystalline systems. Fe$_3$O$_4$ diffraction peaks are observed at 18.0°, 30.2°, 35.5°, 43.3°, 57.2°, 63.0°, and 74.0° referring to reflections (111), (220), (311), (400), (511), (440), and (533), respectively [36,47]. The XRD patterns of the PdSnO$_2$/C binary material are shown in **Fig. 3b**. The XRD analysis further confirmed the SnO$_2$ nanorods structures, with all diffraction peaks being attributed to the tetragonal phase of SnO$_2$, in agreement with reported values (JCPDS nº 77-0450) [43]. The peaks at 26.6°, 34.1°, 38.2°, 51.8°, 54.8°, 58.1°, 61.8°, 64.9°, 66.1°, 71.4°, and 78.9° by reflections (110), (101), (200), (211), (220), (002), (310), (112), (301), (202), and (321). H. Wang *et al*. [43] studied diffraction planes of nanorods and hollow microspheres of SnO$_2$ with pronounced crystallinity. As reported by the authors, the diffraction signals of the SnO$_2$ hollow spheres widened, indicating a smaller particle size than the nanorods. The authors also showed that the intensity of the diffraction peaks of the nanorods was relatively higher, demonstrating enhanced crystallinity and corroborating the diffractogram in which intense and well-defined peaks were observed, referring to the diffraction planes of the SnO$_2$ nanorods. The XRD patterns of the ternary material PdFe$_3$O$_4$SnO$_2$/C are shown in **Fig. 3c**. The reflection planes (220), (311), (400), (511), and (400) of Fe$_3$O$_4$ followed the PDF 19-629 standard so that the diffraction peaks can be correlated to a cubic phase (fcc, space group Fd-3 m) magnetite structure. The reflection planes (110), (101), (211), (220), (002), (310), (301), and (202) of SnO$_2$ confirmed the tetragonal phase of SnO$_2$, in agreement with reported values (JCPDS # 77-0450).

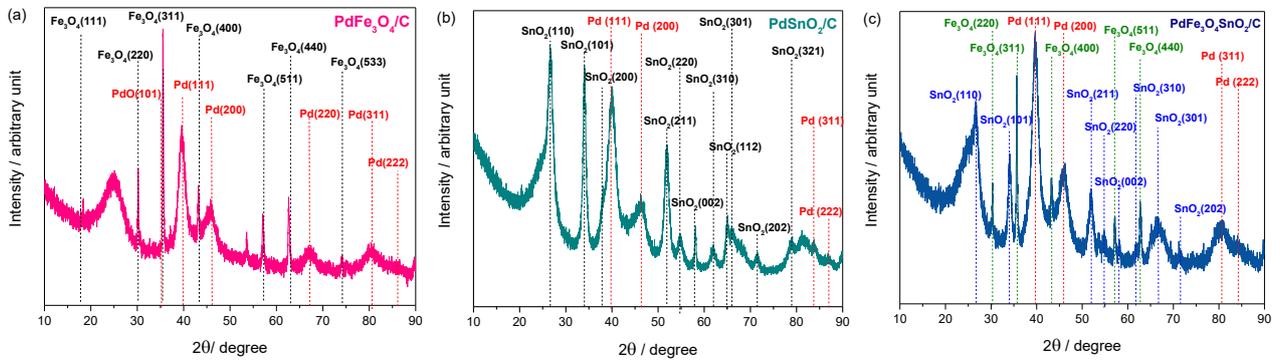

**Fig. 3.** XRD patterns of **(a)** PdFe$_3$O$_4$/C, **(b)** PdSnO$_2$/C, and **(c)** PdFe$_3$O$_4$SnO$_2$/C electrocatalyst.

Based on the diffractograms, the average crystallite size (D) was calculated using the Debye-Scherrer Equation (**Eq. 1**), where ʎ is the X-ray wavelength (1.54056 Å for Cu radiation Kα), β is the width at half height of the peak evaluated in the XRD, K is the constant that depends on the reflection symmetry (K = 1 or 0.94 for spherical particles), and θ is the angle of the XRD in radians [7,8,48,49].

$$D\ hkl = K\ ʎ/\ \beta \cos\theta \qquad (1)$$

The interplanar distance and lattice parameter were also calculated using Bragg's Law. The interplanar distance "d" was obtained with **Eq. 2**, where "n" is the diffraction order, evaluated at 1 for face-centered cubic structures (fcc), as is the case with palladium, Ӄ is the wavelength of the X-ray incident on the sample in nanometers (1.54056 Å for Cu Kα radiation), and θ is the angle of reflection of the incident beam in radians [6].

$$d = \frac{n Ӄ}{2 \operatorname{sen} \theta} \tag{2}$$

Once the value of the interplanar distance (d) was determined, the lattice parameter (a) was calculated using **Eq. 3**, where "h", "k", and "l" are the Miller indices. The lattice parameter is determined by the resultant represented by "a."

$$\frac{1}{d^2} = \frac{h^2 + k^2 + l^2}{a^2} \tag{3}$$

The interplanar distance, lattice parameter, and crystallite size, calculated based on the Pd diffraction peaks (111) and (200), are shown in **Tab. 2**. A shift of 2θ was observed for smaller values and, consequently, an increase in the lattice parameter values of the synthesized materials about the commercial material Pd/C AA and the PDF 5-681 standard [50,51], this may be related to the higher presence of Pd oxides in the catalyst. The PdO can lead to a shift of metallic Pd peaks to smaller angles in XRD patterns, especially in nanostructured materials, because of lattice strains, phase interactions, lattice parameter increase, and crystallite size effects [52], which might induce a change in the Pd d-band, facilitating EOR. In addition, the increase in the oxide content of the materials can be inferred from the rise in binding energy observed in XPS experiments [53,54]. These results will be presented and discussed in subsequent sections. The 2θ shift and the lattice parameter increase might induce a change in the Pd *d*-band, facilitating EOR. Furthermore, it is suggested that due to the magnetic properties of $Fe_3O_4$, the electronic effect may be active in materials with $Fe_3O_4$ nano-octahedra in their composition [35,36,55].

H. Rivera-González *et al*. [36] prepared Pd-$Fe_3O_4$ electrocatalysts, but with nanoparticle-morphology for Pd and $Fe_3O_4$, and observed that the enhanced activity and durability of Pd-$Fe_3O_4$ were associated with a "proximity" between Pd and Fe atoms. Geometric properties such as different crystalline structures, lattice parameters, and particle size can modify the electronic properties and improve the EOR electrocatalytic activity. Even though the material obtained was not an alloy, adding Fe species close to Pd atoms can change the interatomic distance. In addition, different stabilizers used in the synthesis can generate displacements associated with deformations or changes in chemical composition.

Tab. 2. Values of interplanar distance, lattice parameter, and crystallite size.

| Electrocatalyst | Interplanar distance (d)/ nm | | Lattice parameter (a)/ nm | | Crystallite size (XRD) (D)/ nm | Crystallite size (TEM) (D)/ nm |
|---|---|---|---|---|---|---|
| | (111) | (200) | (111) | (200) | | |
| Pd/C AA | 0.2236 | 0.1938 | 0.3872 | 0.3876 | 5.3 ± 0.4 | 5.0 ± 1.2 |
| Pd/C | 0.2287 | 0.1975 | 0.3961 | 0.3949 | 5.2 ± 1.2 | 6.2 ± 1.3 |
| PdFe$_3$O$_4$/C | 0.2275 | 0.1984 | 0.3940 | 0.3967 | 5.1 ± 0.7 | 5.1 ± 0.9 |
| PdSnO$_2$/C | 0.2254 | 0.1956 | 0.3904 | 0.3912 | 4.8 ± 0.1 | 5.4 ± 1.1 |
| PdFe$_3$O$_4$SnO$_2$/C | 0.2266 | 0.1971 | 0.3924 | 0.3943 | 4.7 ± 0.02 | 7.3 ± 1.4 |

One of the proposals of this work was to synthesize nanomaterials with different morphologies and evaluate the possible advantages of such nanostructures. The primary focus was to increase the electrocatalytic activity, as the nanomaterials' size and shape significantly impact their physicochemical properties [56–58]. Thus, the morphology of Fe$_3$O$_4$ nano-octahedra and SnO$_2$ nanorods was evaluated by HR-TEM to confirm the achievement of the nanostructures. **Fig. 4a** shows the HR-TEM image of the unsupported Fe$_3$O$_4$ nano-octahedra, confirming morphology. The average length of the edges was 68.3 ± 14.1 nm, but the presence of nano-octahedra with edges greater than 300 nm was also observed. The hydrothermal route employed was previously described by Wei Lei *et al*. [35] with the formation of octahedral structures exhibiting edge lengths of approximately 500 nm being reported. **Fig. 4b** shows the HR-TEM image of unsupported SnO$_2$ nanorods synthesized by an H$_2$O$_2$-assisted hydrothermal route without surfactants, using SnCl$_2$ as a precursor [43]. The histograms indicate the average length and diameter, which are shown in **Fig. 4e** and **Fig. 4f**, respectively. The nanorods' average length ranged from 44.7 ± 13.2 nm, and the average diameter ranged from 8.2 ± 2.4 nm, characterizing a structure with a high aspect ratio. Nanomaterials with a high aspect ratio, like nanorods, nanotubes, and nanowires, have garnered significant interest because of their distinctive chemical, mechanical, electrical, and optical properties and potential uses in nanodevices [59]. The HR-TEM technique was also performed for all the materials studied to evaluate the nanostructure's anchoring on the carbon support and the distribution of palladium nanoparticles, in addition to EDS monitoring for specific regions of the electrocatalyst surfaces. **Fig. 4c** and **Fig. 4d** show the HR-TEM images of the commercial Pd/C AA and Pd/C, respectively, where the darker points refer to Pd, and the light gray regions refer to carbon. The Pd particle size distribution histogram for the commercial Pd/C AA (**Fig. 4g**) and for the simple synthesized material Pd/C (**Fig. 4h**) were performed based on the HR-TEM images. The average size of the Pd nanoparticles from the commercial Pd/C AA was 5.0 ± 1.2 nm, and the average size of the Pd nanoparticles supported on the Vulcan XC-72 carbon for the Pd/C material was 6.2 ± 1.3 nm, like the average crystallite sizes determined by XRD.

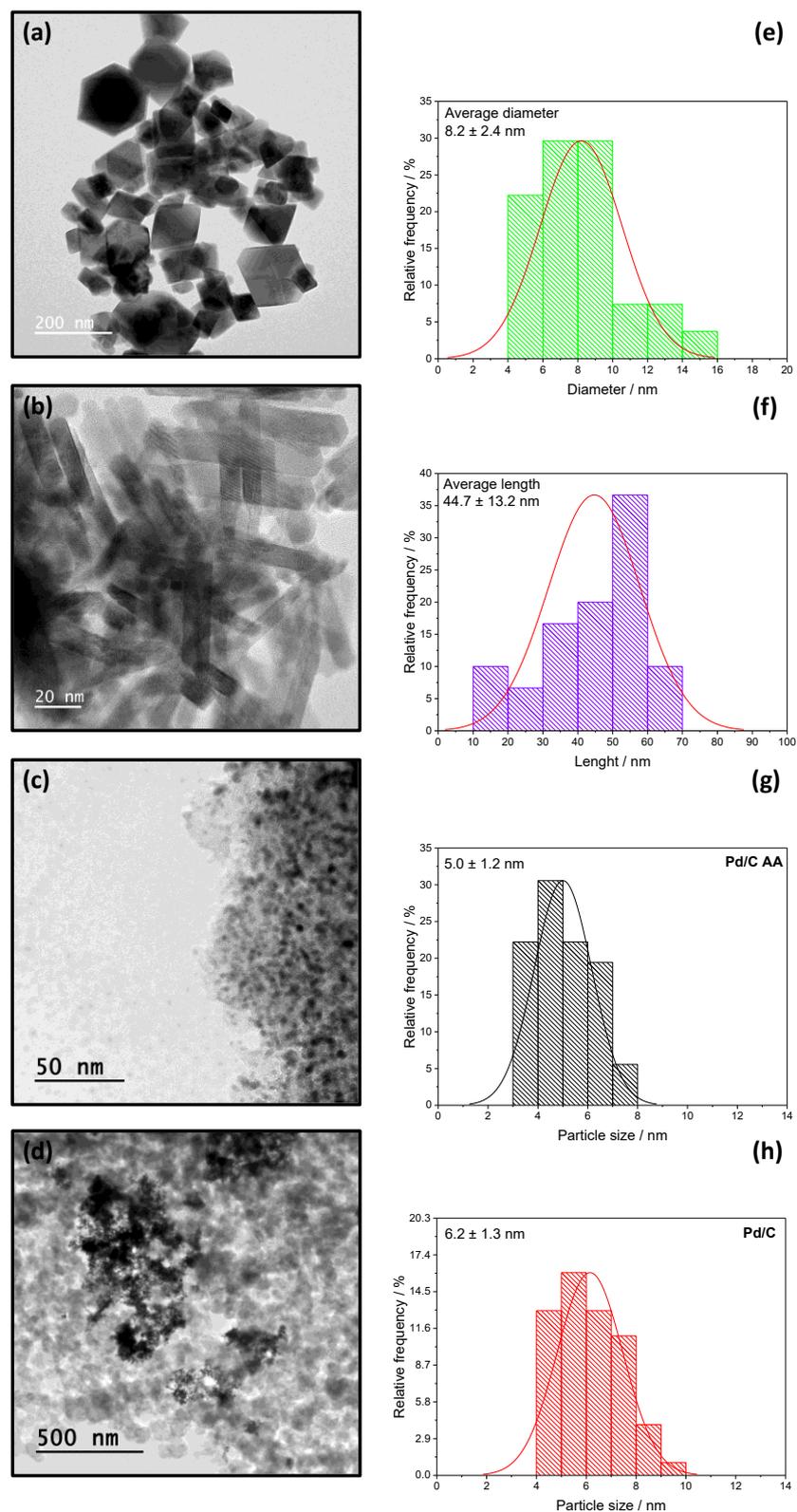

**Fig. 4.** HR-TEM image of **(a)** unsupported $Fe_3O_4$ nano-octahedra, **(b)** unsupported $SnO_2$ nanorods, **(c)** Pd/C AA, and **(d)** Pd/C, **(e)** Average diameter histogram of $SnO_2$ nanorods, **(f)** Average length histogram of $SnO_2$ nanorods. Histogram of the particle size distribution for **(g)** Pd/C AA electrocatalyst and **(h)** Pd/C.

The HR-TEM images and Pd particle size distribution histograms for the $PdFe_3O_4$/C electrocatalyst and EDS were also analyzed, as shown in **Fig. S3a**. The average size of Pd nanoparticles was 5.1 ± 0.9 nm, according to the crystallite size calculated by XRD. The HR-TEM images of the PdSnO$_2$/C material shown in **Fig. S3b** demonstrate the presence of Pd nanoparticles and SnO$_2$ nanorods, and according to the histogram, the Pd particle size was 5.4 ± 1.1 nm. It was possible to verify the anchorage of both nanostructures on the support, in addition to showing the proximity of the SnO$_2$ nanorods to the Pd, which may facilitate the supply of oxygenated species from the oxide, favoring electrocatalytic activity. The HR-TEM images of the ternary material PdFe$_3$O$_4$SnO$_2$/C shown in **Fig. S3c** indicated a less homogeneous surface than the binary materials. The characteristic regions of Pd and Fe$_3$O$_4$ agglomerates could be observed, while the SnO$_2$ nanorods were less agglomerated on the Vulcan XC-72 support. The average particle size was 7.3 ± 1.4 nm, slightly higher than that observed in the other two synthesized binary materials.

Raman spectra were recorded in the range of 1200-1700 cm$^{-1}$ to investigate the structural modifications of Vulcan XC-72 carbon following its modification with Pd nanoparticles, Fe$_3$O$_4$ nano-octahedra, and SnO$_2$ nanorods in the simple, binary, and ternary electrocatalyst. Raman spectroscopy is considered sensitive to the carbon structure and interstitial changes that disturb the translational symmetry of the analyzed sample since Raman scattering refers to an inelastic behavior during the interaction of light with matter and is intrinsic to each material [60,61].

Vulcan XC-72 carbon (carbon black) predominantly consists of sp²-hybridized carbon planes, and its Raman spectrum exhibits characteristic bands at approximately 1350 cm$^{-1}$ (D-band) and 1580 cm$^{-1}$ (G-band). The D-band is due to the respiration vibration of the hexagonal carbon lattice and is related to structure disorder, which are defects of the functional group containing oxygen on the carbon surface, and the G-band refers to the elongation of carbon-carbon bonds, resulting from vibrations in the plane of sp$^2$ carbons. To evaluate the defects in the material, a ratio is made between the areas of the Raman $I_D/I_G$ shift bands, being able to infer a more significant number of defects at a high $I_D/I_G$ ratio and a smaller number of defects at a low ratio [62].

**Fig. S4a** shows the Raman spectra of the Vulcan XC-72 carbon and the Pd/C simple electrocatalyst, indicating the D-band peaks at 1350 cm$^{-1}$ and the G-band peaks at 1580 cm$^{-1}$ of carbon black. **Fig. S4b** shows the Raman spectra of the PdFe$_3$O$_4$/C, PdSnO$_2$/C, and PdFe$_3$O$_4$SnO$_2$/C, indicating the D-band and G-band. $I_D/I_G$ ratios were calculated as shown in **Tab. 3**, and it was verified that all materials containing Pd nanoparticles and nanostructures presented with an $I_D/I_G$ ratio more significant than Vulcan XC-72 carbon due to defects caused in the sp$^2$ carbon structure. The same effect was observed for the commercial material Pd/C AA, as published by Pinheiro *et al*. [4]. The authors state that the defects may be directly related to favoring the electrocatalytic performance of the material and refer, in this case, to the state of carbon oxidation. It is suggested that higher amounts of defects and oxygen-containing species in the sp$^2$ carbon structure may improve electronic transport. Their results indicate that the modifications on the carbon surface promoted by metals could contribute to electrocatalytic activity.

**Tab. 3.** $I_D/I_G$ ratios of the Raman spectra of Vulcan XC-72 carbon, Pd/C, PdFe$_3$O$_4$/C, PdSnO$_2$/C, and PdFe$_3$O$_4$SnO$_2$/C electrocatalysts.

| Material | Ratio $I_D/I_G$ |
|---|---|
| C Vulcan XC-72 | 0.78 ± 0.17 |
| Pd/C | 1.24 ± 0.17 |
| PdFe$_3$O$_4$/C | 0.88 ± 0.23 |
| PdSnO$_2$/C | 0.85 ± 0.04 |
| PdFe$_3$O$_4$SnO$_2$/C | 0.93 ± 0.09 |

**Fig. 5** shows the high-resolution XPS spectra of C 1s, the Pd/C (a), PdFe$_3$O$_4$/C (b), PdSnO$_2$/C (c), and PdFe$_3$O$_4$SnO$_2$/C (d) samples. The high-resolution spectra of Pd 3d, Fe 2p, and Sn 3d for the respective electrocatalysts are shown in **Fig. S5**. The C 1s spectra were fitted using 4 components located at binding energies of approximately 284 eV, 285, 286, and 289 eV corresponding to the C–C, C–OH, C=O, and COOH bonds, respectively [63–65]. The PdFe$_3$O$_4$/C, PdSnO$_2$/C, and PdFe$_3$O$_4$SnO$_2$/C electrocatalysts presented a percentage of oxygenated groups of 52.37, 51.04, and 49%, respectively. While the Pd/C electrocatalyst presented a lower amount of oxygenated groups of 45.36%. The greater amount of oxygenated groups can improve the electrocatalytic activity of the material, with the PdFe$_3$O$_4$/C electrocatalyst being the one with the highest amount of oxygenated groups [4].

The Pd 3d spectra were deconvoluted for all electrocatalysts using two components related to Pd$^0$ and Pd$^{2+}$ located at approximately 335 eV and 336 eV, respectively [66]. The Pd/C electrocatalyst presented 46.89% of Pd oxide, while the PdFe$_3$O$_4$/C, PdSnO$_2$/C, and PdFe$_3$O$_4$SnO$_2$/C electrocatalysts presented 59.36, 57.44, and 54.07%, respectively. The increase in the percentage of Pd oxides may be associated with the modification of the materials by Fe and Sn oxides, and the presence of Pd oxides in the electrocatalysts can improve the electrocatalytic properties by providing oxygen species necessary for the EOR oxidation intermediates [4,67]. Furthermore, it can be observed from **Fig. S5** that the $3d_{5/2}$ and $3d_{3/2}$ peaks of Pd for the synthesized materials PdFe$_3$O$_4$/C, PdSnO$_2$/C, and PdFe$_3$O$_4$SnO$_2$/C were shifted to higher energies by about 0.5 eV compared to Pd/C, indicating a loss of electron density, which corroborates the XRD results shown previously, which indicated a greater presence of Pd oxides in the binary and ternary materials [53,54,67]. This shift is indicative of electronic interactions between Pd and Fe and Sn arising from the metal oxides (Fe$_3$O$_4$ and SnO$_2$), suggesting the withdrawal of electrons from Pd due to the strong metal–oxide interaction. The presence of Fe and Sn modifies the electronic structure of Pd, leading to a downward shift of the d-band center of Pd. This electronic modulation weakens the Pd-adsorbate bond (e.g., for CO and CH$_3$CH$_2$OH intermediates), facilitating their desorption and improving the tolerance of the electrocatalyst to toxic species. As a result, these electronic effects, combined with possible bifunctional mechanisms (e.g., OH$_{ads}$ provided by Fe$_3$O$_4$ or SnO$_2$), synergistically contribute to the enhanced electrocatalytic performance of binary and ternary catalysts in the EOR [53,54].

The Sn 3d spectra of PdSnO$_2$/C and PdFe$_3$O$_4$SnO$_2$/C electrocatalysts showed two peaks located at 495 eV (Sn$^{4+}$) and 487.0 eV (Sn$^{4+}$), which indicate the presence of SnO$_2$ and confirm the presence of Sn in the electrocatalysts [68]. The Fe 2p spectra of PdFe$_3$O$_4$/C and PdFe$_3$O$_4$SnO$_2$/C electrocatalysts were deconvoluted using two components related to Fe$^{2+}$ and Fe$^{3+}$ located at approximately 711 and 725 eV, respectively, confirming the presence of Fe in the electrocatalysts [69].

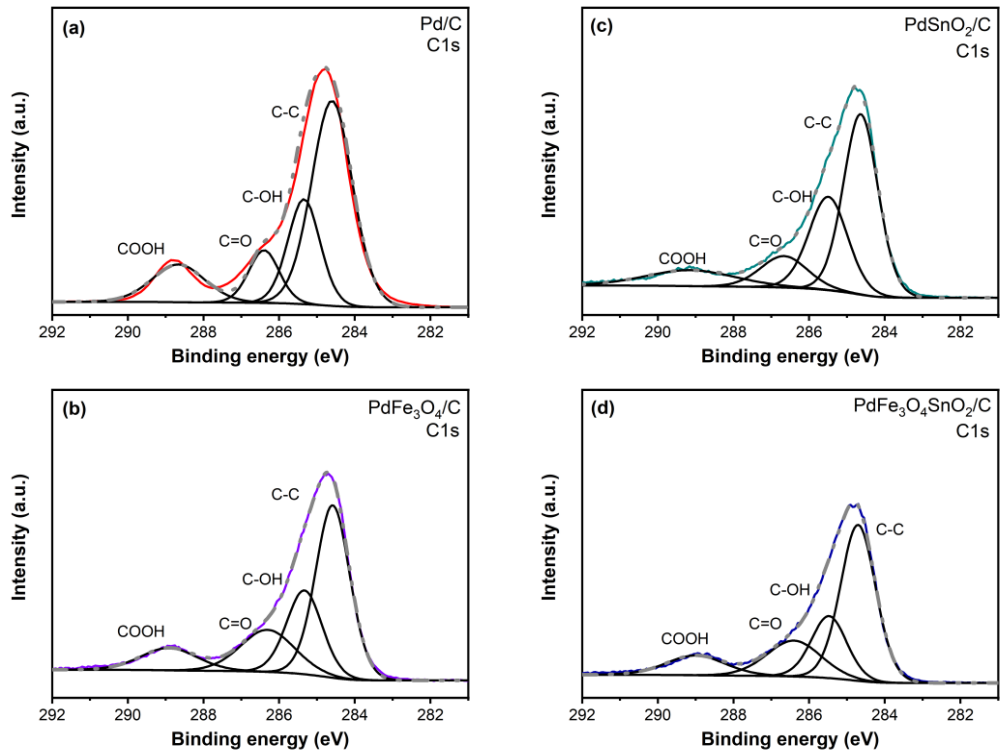

**Fig. 5.** Fitted XPS high-resolution core-level spectra of C1s (a) Pd/C, (b) PdFe$_3$O$_4$/C, (c) PdSnO$_2$/C, and (d) PdFe$_3$O$_4$SnO$_2$/C.

**Electrochemically Active Surface Area (ECSA)**

CO-stripping experiments were performed to evaluate the ECSA and the operation of bifunctional systems; CO is one of the main species responsible for catalytic poisoning in oxidation processes of small organic molecules and one of the challenges in the operation of DLFC [6,16].

The literature discusses that the removal or CO-stripping occurs by the reaction between OH$_{ads}$ and CO$_{ads}$ formed (**Eq. 4**); the CO is adsorbed on the palladium active sites (Pd-CO$_{ads}$) [6,16].

$$CO_{ads} + OH_{ads} \longrightarrow CO_2 + H_2O \qquad (4)$$

The ECSA was evaluated through the electrochemical oxidation of a previously adsorbed CO layer, and calculated using **Eq. 5**:

$$ECSA = \frac{Q}{G \times 420} \qquad (5)$$

Here, Q denotes the desorption-electrooxidation charge of CO in microcoulombs (μC), G represents the mass in grams (g) of metal (Pd or Pt), and 420 refers to the charge required to oxidize a monolayer of CO adsorbed on the catalyst in microcoulomb per centimeter square ($\mu C\ cm^{-2}$) [6,16].

**Fig. S6** shows the voltammograms for CO-stripping performed in alkaline medium using electrocatalysts based on Pd nanoparticles. The ECSA of the electrocatalysts in $m^2 g^{-1}$ Pd was estimated by CO oxidation. The start potential ($E_{onset}$) for the electrochemical oxidation of the pre-adsorbed CO layer was displayed in **Tab. 4**. The increasing order of ECSA was PdC AA < PdSnO$_2$/C < Pd/C < PdFe$_3$O$_4$SnO$_2$/C < PdFe$_3$O$_4$/C, and the electrocatalysts containing Fe$_3$O$_4$ nano-octahedra and SnO$_2$ nanorods showed a less positive $E_{onset}$ than the simple materials. The PdFe$_3$O$_4$/C electrocatalyst stood out among the others, followed by the PdFe$_3$O$_4$SnO$_2$/C material, suggesting that the presence of Fe$_3$O$_4$ nano-octahedra and SnO$_2$ nanorods may favor the bifunctional mechanism, facilitating the oxidation of CO to CO$_2$ and, consequently, reducing catalytic poisoning.

**Tab. 4.** ECSA of electrocatalysts by analysis of CO-stripping and $E_{onset}$ of CO removal.

| Electrocatalyst | ECSA ($m^2\ g^{-1}$ Pd) | $E_{onset}$ (V vs. SCE) |
|---|---|---|
| PdC AA | 6.7 | -0.311 |
| Pd/C | 21.9 | -0.343 |
| PdFe$_3$O$_4$/C | 36.5 | -0.494 |
| PdSnO$_2$/C | 10.8 | -0.447 |
| PdFe$_3$O$_4$SnO$_2$/C | 31.8 | -0.432 |

**Study of electrochemical characterization and EOR in alkaline medium**

In electrochemical characterization processes, three peaks are observed in the positive sweep of a voltammetric profile, corresponding to different phenomena occurring on the Pd electrode's surface. Studies by Z.X. Liang *et al.* [70] defined these regions under similar conditions to those studied in this work. The authors showed the first peak is due to the oxidation of absorbed and adsorbed hydrogen (**Eq. 6**).

$$Pd\text{-}H_{abs/ads} + OH^- \longrightarrow Pd + H_2O + e^- \quad (6)$$

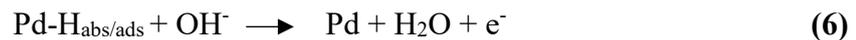

OH$^-$ adsorption (**Eq. 7**) starts at the negative potential far from the Pd onset potential (**Eq. 8 and 9**) and partially overlaps with the hydrogen desorption peak, generating a second one. A third peak can be attributed to forming the palladium (II) oxide layer on the catalyst surface. It has been widely accepted that OH$^-$ ions are first chemisorbed in the early stage of oxide formation [70].

$$Pd + OH^- \longleftrightarrow Pd\text{-}OH_{ads} + e^- \quad (7)$$

$$Pd\text{-}OH_{ads} + OH^- \longleftrightarrow PdO + H_2O + e^- \quad (8)$$

$$Pd\text{-}OH_{ads} + Pd\text{-}OH_{ads} \longleftrightarrow PdO + H_2O \quad (9)$$

A fourth peak appears corresponding to the oxidation process, attributed to the reduction of Pd(II) oxide during the reverse scan (**Eq. 10**) [70].

$$PdO + H_2O + 2e^- \longleftrightarrow Pd + 2OH^- \quad (10)$$

**Fig. S7** shows the electrochemical characterization profiles in absence of ethanol by cyclic voltammetry for the Pd/C AA, Pd/C, PdFe$_3$O$_4$/C, PdSnO$_2$/C, and PdFe$_3$O$_4$SnO$_2$/C electrocatalysts. All electrocatalysts showed characteristic profiles for Pd as regions of hydrogen adsorption/desorption and the PdO formation and reduction process. The materials modified with Fe$_3$O$_4$ and SnO$_2$ showed a broadening in the electrical double-layer region compared to the simple Pd/C material; this effect may be associated with oxygenated species on the Pd surface. The results obtained in this work corroborate previous studies by H. An *et al.* [71] in which electrocatalysts modified with SnO$_2$ exhibited a greater ECSA and superior electrocatalytic performance for alcohol oxidation.

**Fig. 6a** shows the cyclic voltammetry for EOR in alkaline medium catalyzed by the materials Pd/C AA, Pd/C, PdFe$_3$O$_4$/C, PdSnO$_2$/C, and PdFe$_3$O$_4$SnO$_2$/C, using 1.0 mol L$^{-1}$ ethanol in 1.0 mol L$^{-1}$ KOH, with a scan rate of 20 mV s$^{-1}$ and a potential window from -1.0 to +0.2 V vs. SCE. It is observed that the hydrogen adsorption/desorption region was suppressed in the presence of ethanol in the solution. A current peak centered around -0.1 V vs. SCE is observed during the anodic sweep, and on the cathodic sweep, another current peak is found centered around -0.3 V vs. SCE. In the hydrogen region, the peak suppression can be attributed to the ethanol dissociative adsorption in the low potential region. The resulting ethoxy, such as (CH$_3$CO)$_{ads}$, are strongly adsorbed on the active sites of the Pd electrode, blocking the hydrogen adsorption and reducing hydrogen peaks, according to **Eq. 11** and **12** [70].

$$Pd + CH_3CH_2OH \longleftrightarrow Pd\text{-}(CH_3CH_2OH)_{ads} \quad (11)$$

$$Pd\text{-}(CH_3CH_2OH)_{ads} + 3OH^- \longrightarrow Pd\text{-}(CH_3)_{ads} + 3H_2O + 3e^- \quad (12)$$

The oxidation current can only be observed when the potential is above the initial potential of the hydroxyl ions adsorption, indicating that the adsorbed intermediates formed during the dissociative adsorption of ethanol can be removed from the Pd electrode by the adsorbed species containing oxygen (Pd-OH$_{ads}$). As a result, EOR can progress continuously, causing the current to increase with potential, as shown in CV. After the anodic oxidation peak reaches its maximum value, a decrease in current is observed with growing potential, which may be related to the PdO formation on the electrode surface under higher potential conditions, as detailed in the equations of the electrochemical characterization experiments in the absence of fuel. The development of oxide layers can obstruct the adsorption of reactive species on the Pd surface and decrease electrocatalytic activity. The EOR oxidation current decreases with the increase in potential in the forward scan due to the more significant amount of PdO formed on the electrode surface, reaching a stage in which the oxidation current practically coincides with the current in the electrolytic support solution, indicating that the EOR that occurs in the oxide layer is insignificant. Electrocatalytic activity is recovered during the negative scan, as evidenced by the peak current shown in the CV, and this reactivation can be attributed to the reduction of Pd(II) oxide. The "overshoot" in current during the reverse scan is due to the accumulation near the surface of the reagent, which cannot react when the surface is oxidized. Therefore, when the oxide is reduced, an overpopulation of reagents can react, leading to a sharp increase in current [70].

From the cyclic voltammograms, the results of mass activity for EOR were evaluated, as shown in **Tab. 5**. The PdFe$_3$O$_4$/C demonstrated outstanding performance in the EOR experiments, despite containing approximately 45% less Pd compared to commercial Pd/C AA. It exhibited the highest mass activity among the studied catalysts (1426 mA mg$^{-1}$Pd), nearly twice that of the commercial and simple Pd/C, and was followed by PdSnO$_2$/C (1135 mA mg$^{-1}$Pd). This enhanced performance may be attributed to the presence of Fe$_3$O$_4$ nano-octahedra and SnO$_2$ nanorods, which can promote the bifunctional mechanism [27,29,72] by supplying oxygenated species [4,73–75]. The increasing order of mass activity was Pd/C< Pd/C AA< PdFe$_3$O$_4$SnO$_2$/C< PdSnO$_2$/C< PdFe$_3$O$_4$/C. Regarding the onset potential (E$_{onset}$), both the binary and ternary materials containing SnO$_2$ nanorods resulted in less positive values than the others and improved catalytic activity compared to the Pd/C [76].

The results presented in this work show that the modification of low-Pd amount electrocatalysts using nanostructured metal oxide cocatalysts, such as Fe$_3$O$_4$ nano-octahedra and SnO$_2$ nanorods, can significantly increase the electrooxidation of ethanol in alkaline media. The oxides can contribute different functionalities that complement the activity of Pd. In the case of $Fe_3O_4$, [35] its nano-octahedral morphology exposes mostly (111) facets, as observed in XRD tests, which can favor the adsorption of hydroxyl — an essential step in the activation of ethanol and in the oxidative removal of intermediates that cause catalytic poisoning, such as $CO_2$ [21,25,77].

In the case of $SnO_2$ nanorod morphology [43], with its high aspect ratio and high surface area, it facilitates the supply of oxygenated species through a bifunctional mechanism, resulting in improved removal of adsorbed intermediates. Furthermore, the proximity between these nanostructures and Pd can improve nanoparticle dispersion and long-term stability. The aforementioned morphological effects help to explain the observed improvements in both EOR activity and durability when Pd/C is combined with these oxide nanostructures [25,35,43].

There is no evidence in the literature indicating that isolated $SnO_2$/C exhibits significant electrocatalytic activity toward ethanol oxidation [37]. Studies have shown that $SnO_2$ can contribute to ethanol oxidation by facilitating the adsorption of hydroxyl species ($OH^-$), which aids in the removal of intermediates such as CO. Additionally, $SnO_2$ commonly serves as a conductive and structural support for metal-based electrocatalysts [37,78,79]. However, its intrinsic catalytic activity is not sufficient for practical applications when used alone. $Fe_3O_4$ exhibits negligible direct activity toward EOR and acts primarily as a cocatalyst, improving performance through bifunctional effects or electronic interactions when combined with active noble metals [36,47,80]. The current in this study was normalized by the amount of Pd used, since Pd is the active site responsible for ethanol oxidation. Furthermore, considering the high cost of Pd, reporting the activity per mass of Pd (mass activity) provides important information to assess the feasibility of using the catalyst and the cost-performance balance, widely adopted in electrocatalysis research [56,66,81–83].

**Tab. 5**. Onset potential, mass activity results for EOR and chronoamperometric current using the Pd/C AA, Pd/C, PdFe$_3$O$_4$/C, PdSnO$_2$/C, and PdFe$_3$O$_4$SnO$_2$/C electrocatalysts.

| Electrocatalyst | $E_{onset}$/ mV vs. SCE | Mass activity/ mA mg$^{-1}$Pd | Current to EOR after 1800 s/ mA mg$^{-1}$Pd |
|---|---|---|---|
| Pd/C AA | -799 | 651 | 149 |
| Pd/C | -792 | 597 | 160 |
| PdFe$_3$O$_4$C | -813 | 1426 | 512 |
| PdSnO$_2$C | -805 | 1135 | 447 |
| PdFe$_3$O$_4$SnO$_2$C | -811 | 1074 | 439 |

The behaviors for EOR in alkaline medium on the electrocatalysts' surfaces were studied using chronoamperometry (CA) measurements at -0.3 V vs. SCE for 1800 seconds. **Fig. 6b** shows these results of chronoamperometric measurements for Pd/C AA, Pd/C, PdFe$_3$O$_4$/C, PdSnO$_2$/C, and PdFe$_3$O$_4$SnO$_2$/C, using 1.0 mol L$^{-1}$ ethanol in KOH 1.0 mol L$^{-1}$. In CA, higher initial currents are observed, followed by a decrease due to polarization. This occurs due to the electrical double-layer charging in the electrode, which goes on until it reaches equilibrium. In addition, the first minutes of chronoamperometry show a current decline, indicating a more significant contribution of diffusion overpotential. Subsequently, a pseudo-steady state is reached due to the inactivation of catalytic sites by reaction intermediates (CO, CH$_x$, CH$_3$CHO) [18,84,85]. The PdFe$_3$O$_4$/C material for EOR generated a higher current density than the other materials, resulting in the value of 512 mA mg$^{-1}$ Pd (**Tab. 5**), followed by the PdSnO$_2$C, PdFe$_3$O$_4$SnO$_2$C, Pd/C, and Pd/C AA. The significant decrease in the electrocatalytic activity of the simple materials Pd/C and Pd/C AA may be associated with catalytic poisoning caused by strongly adsorbed intermediates on palladium active sites. This can be observed in the oxidation currents at the end of chronoamperometry, at 1800 seconds. It is suggested that the Fe$_3$O$_4$ nano-octahedra and the SnO$_2$ nanorods may have minimized the effects of catalytic poisoning, facilitating the oxidation of intermediates by providing oxygen-containing species, thus releasing the Pd active sites since the CA currents were higher for the binary and ternary materials [6,37,86–89]. Both effects favor the oxidation of the intermediates adsorbed on the surface of the electrocatalyst, releasing the active sites for the oxidation of a new fuel molecule [4,90–92].

The increase in the number of defects on the surface of the electrocatalyst due to the presence of nanostructures can also favor the electrocatalytic activity, corroborating the Raman results, in which a higher I$_D$/I$_G$ ratio was observed for the modified materials, indicating a more significant amount of oxygenates species, vacancies and defects in the carbon structure [4]. Furthermore, these results were consistent with the CV measurements for EOR mentioned above.

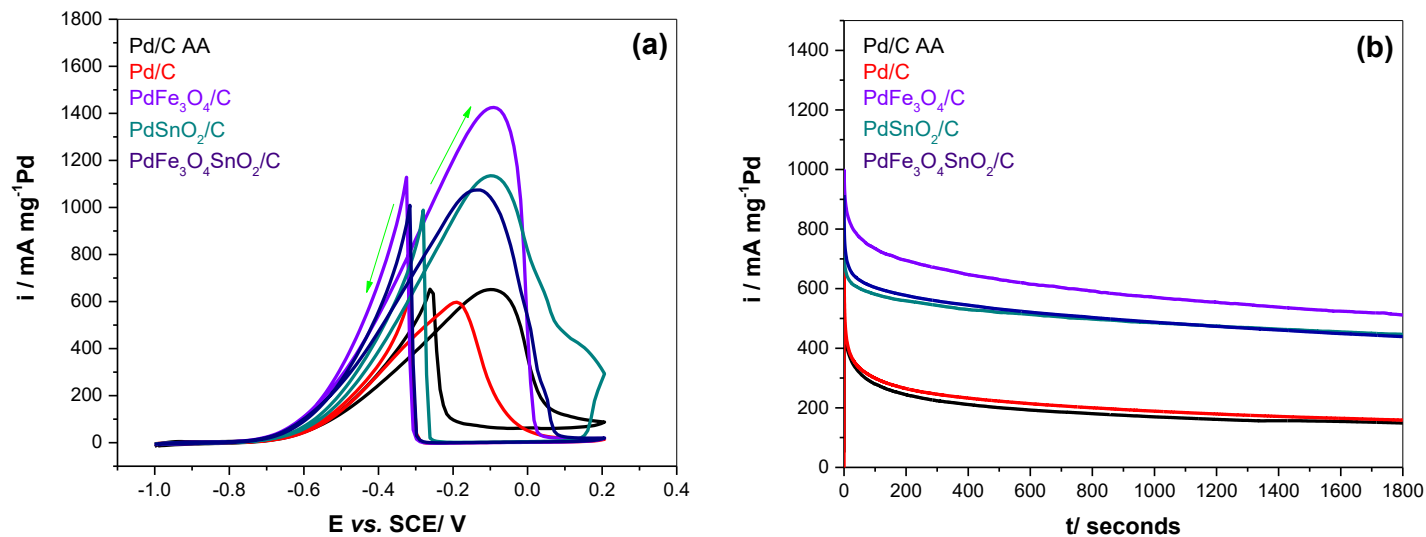

**Fig. 6. (a)** Cyclic voltammetry performed in the presence of 1.0 mol L$^{-1}$ ethanol in 1.0 mol L$^{-1}$ KOH in the potential window -1.0 V and 0.2 V vs. SCE, with a scan rate of 20 mV s$^{-1}$. **(b)** Chronoamperometry obtained for the electrocatalysts Pd/C AA, Pd/C, PdFe$_3$O$_4$/C, PdSnO$_2$/C, and PdFe$_3$O$_4$SnO$_2$/C, using 1.0 mol L$^{-1}$ ethanol in KOH 1.0 mol L$^{-1}$, applying - 0.3V vs. SCE.

**Fig. S8.** shows the Tafel plot of EOR, the Tafel slope, and the exchange current density ($i_0$) for EOR catalyzed by materials. The EOR Tafel plot was performed close to the steady state using linear scanning voltammetry (LSV) at room temperature, and a sweep speed of 0.5 mV s$^{-1}$. In the LSV experiments, it was possible to get the Tafel slope, defined by the mathematical relation shown in **Eq. 13**:

$$\eta = \frac{-2.303\,RT}{\beta F}\log i_0 + \frac{2.303\,RT}{\beta F}\log i \qquad (13)$$

Where $\eta$ is the overpotential, $\beta$ is the anodic transfer coefficient, $i_0$ is the exchange current density, i is the current density, R is the universal gas constant, F is the Faraday constant, and T is the temperature [29,70].

According to the EOR Tafel plot, the PdFe$_3$O$_4$/C electrocatalyst exhibited the highest value of $i_0$, indicating a higher electron transfer rate than the other electrocatalysts [93,94]. Some authors suggest that a lower Tafel coefficient is related to greater efficiency for EOR at lower potential. However, EOR is a multielectron reaction, and the Tafel coefficient is associated with a simple electrochemical process of electron transfer, where the number of electrons is determined by the rate of the slower stage, not reflecting a complex EOR system [94,95].

Currently, the most accepted mechanism for the EOR on the Pd surface in alkaline media can be expressed by **Eq. 14-17**. Due to easier adsorption of hydroxyl on Pd, the carbonaceous species are removed, and an increase in current can be observed. To preserve the adsorbed OH⁻ species on the surface, it is necessary to weakly adsorb water molecules on the Pd sites that will dissociate into $OH^-_{ads}$ and $H^+$. The adsorbed OH- species on the catalytic surface facilitate the ethanol dehydrogenation reaction. The mass activity can be attributed to the presence of oxophilic species that improve the electrocatalyst's ability to remove adsorbed CO on the Pd active sites and adsorb hydroxyl, increasing the mass activity of the EOR [29,96–98].

$$Pd + OH^- \rightarrow Pd\text{-}OH_{ads} + e^- \quad (14)$$

$$Pd + CH_3CH_2OH + 3\ OH^- \rightarrow Pd\text{-}(CH_3CO)_{ads} + 3\ H_2O + 3\ e^- \quad (15)$$

$$Pd\text{-}(CH_3CO)_{ads} + Pd\text{-}OH_{ads} \rightarrow Pd + CH_3COOH + Pd \quad (16)$$

$$Pd\text{-}CH_3COOH + OH^- \rightarrow Pd + CH_3COO^- + H_2O \quad (17)$$

**Online SFC-ICP-MS - Potential-Dependent Dissolution Rates of Pd-based materials in alkaline medium**

A stability study was performed for the most promising material according to the electrochemical studies; it was then compared to the simple Pd-based catalyst using an online SFC-ICP-MS. A potential range was established to examine the Pd dissolution mechanism to measure at which potential Pd dissolution is expected, mapping the stability of the Pd/C and PdFe₃O₄/C materials in the presence of ethanol. The protocol used showed CVs were registered using + 0.3 V$_{RHE}$, +0.5 V$_{RHE}$, +0.7 V$_{RHE}$, +0.9 V$_{RHE}$, +1.0 V$_{RHE}$, and +1.3 V$_{RHE}$ upper potential limits at a scan rate of 5 mV s⁻¹ and a last hold applying +0.77 V$_{RHE}$ for 600 s. The dissolution mappings were recorded in the presence of ethanol using 0.05 mol L⁻¹ KOH as the electrolyte (**Fig. 7**) (0.05 mol L⁻¹ KOH and 0.05 mol L⁻¹ EtOH solution), with transient Pd dissolution only beginning during the forward scan from the fourth CV at the +0.9 V$_{RHE}$ upper potential limit (UPL). The contact peak (which emerged when the SFC contacted the working electrode) was more pronounced in the Pd/C than in the PdFe₃O₄/C, and no contact peak or dissolution for Fe was observed, which was expected in an alkaline medium. Therefore, we suggest that iron contributes to the enhanced stability of the electrocatalyst, as PdFe₃O₄ exhibits reduced Pd dissolution, indicating a stabilizing effect associated with the presence of FeO₄ [36,47].

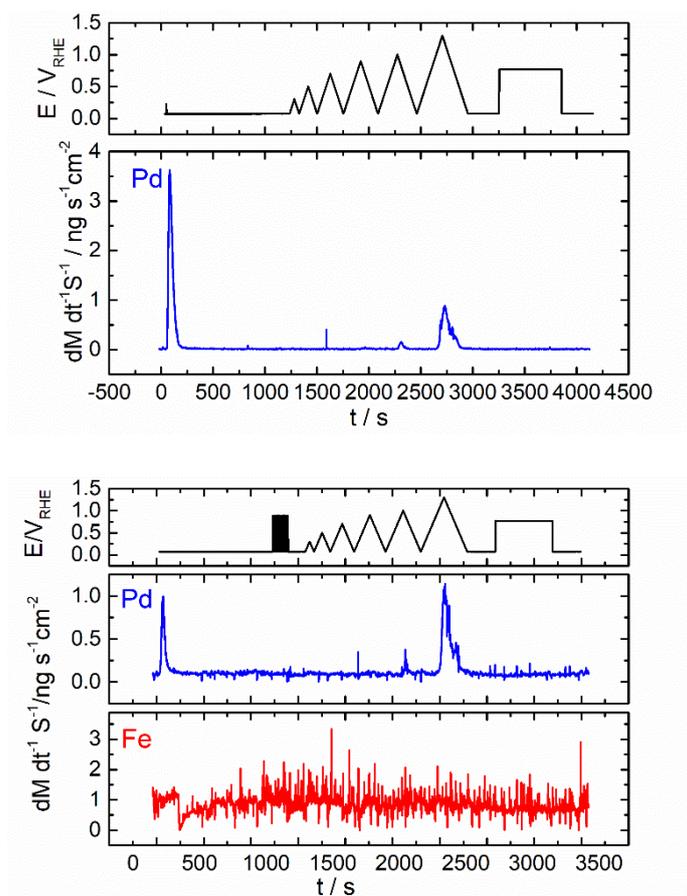

**Fig. 7.** Dissolution rates of Pd/C and PdFe$_3$O$_4$/C electrocatalysts recorded during the electrochemical protocol in a 0.05 mol L$^{-1}$ KOH and 0.05 mol L$^{-1}$ EtOH solution.

**Experiments in Alkaline Direct Ethanol Fuel Cell (ADEFC)**

The Pd/CAA, Pd/C, PdFe$_3$O$_4$/C, PdSnO$_2$/C, and PdFe$_3$O$_4$SnO$_2$/C electrocatalysts were applied in an ADEFC as anode to catalyze EOR, and the commercial Pt/C Alfa Aesar was used as cathode in all experiments. The polarization and power density curves for each material are shown on **Fig. 8**. Using the anode electrocatalyst PdFe$_3$O$_4$/C, the highest open-circuit voltage (OCV) was obtained, exhibiting a value of 1107 mV at an operating temperature of 70 °C. This corroborates the Tafel results, in which this material showed the highest value of i$_0$, possibly indicating a higher electronic transfer rate and a Enhanced electrocatalytic activity.

The electrocatalysts PdFe$_3$O$_4$/C, PdSnO$_2$/C, and PdFe$_3$O$_4$SnO$_2$/C stood out among the others in terms of power density, mainly at the operating temperature of 70 °C, where the values obtained were 31 mW cm$^{-2}$, 30 mW cm$^{-2}$, and 28 mW cm$^{-2}$, respectively. These results corroborate previous electrochemical studies in which materials containing Fe$_3$O$_4$ nano-octahedra and SnO$_2$ nanorods resulted in higher EOR CA currents and higher ECSA than those obtained with simple materials, suggesting greater tolerance to catalytic poisoning. The increase in the number of defects on the material surface modified with nanostructures according to Raman results and the presence of oxides, as confirmed by XRD, EDS, and HR-TEM, may have favored the bifunctional mechanism due to the supply of oxygenated species, and electronic effect due to the magnetic properties of $Fe_3O_4$ and due to the presence of nanostructures, also corroborating the operation results. **Tab. S1** shows all OCV, current density, and power density results obtained in the ADEFC operation studies.

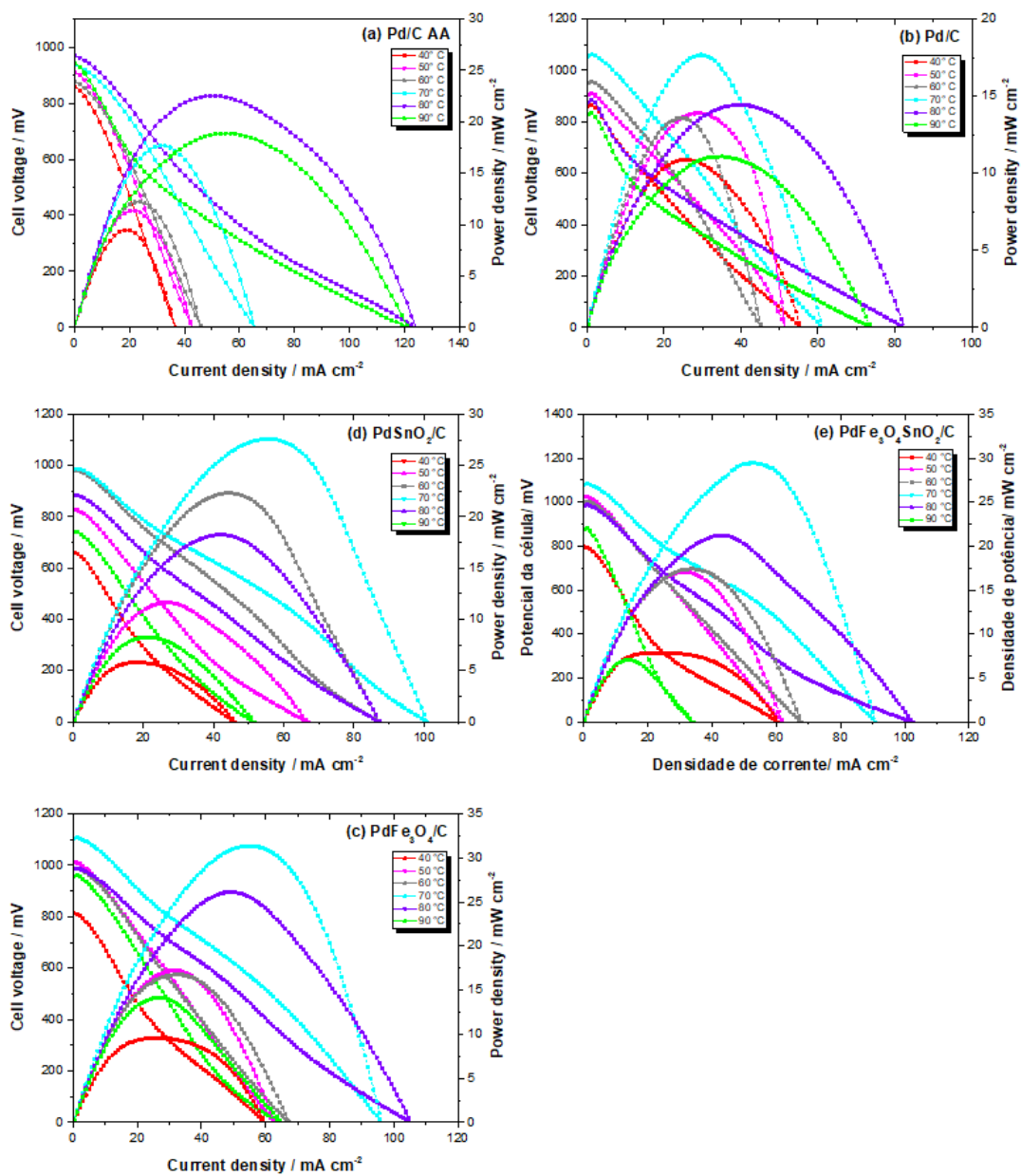

**Fig. 8.** Polarization and power density curves in a unitary ADEFC, using 2.0 mol L$^{-1}$ ethanol in 1.0 mol L$^{-1}$ KOH as fuel, at 1.0 mL min$^{-1}$ anode flow and oxygen humidifier set at 85 °C with 200 mL min$^{-1}$

cathode flow, using commercial cathode electrocatalyst Pt/C Alfa Aesar and anode electrocatalysts (a) Pd/C AA, (b) Pd/C, (c) PdFe$_3$O$_4$/C, (d) PdSnO$_2$/C, and (e) PdFe$_3$O$_4$SnO$_2$/C, at different operating temperatures.

The operating results obtained in this work were compared with the literature (**Tab. 6**) under similar conditions in terms of flow, operating temperature, fuel concentration, and electrolyte concentration. Superior results to those obtained in this work have been reported in the literature, but with higher Pd amounts (by a factor of 2-4), implying a costly material. The main objective of this research is to propose replacing part of the Pd metallic amount with metallic oxides to reduce the cost and improve the electrocatalytic activity of novel electrocatalysts [25,26,99,100]. Promising results were obtained for synthesized electrocatalysts in this work containing 0.5 mg Pd cm$^{-2}$, surpassing previous results from the literature with higher Pd amounts (1.0 mg Pd cm$^{-2}$) under similar operating conditions.

**Tab. 6.** Comparison with the performance of electrocatalysts in ADEFC reported in the literature.

| Anode | Operation conditions | Noble metal anodic amount (mg Pd cm$^{-2}$) | Maximum power density (mW cm$^{-2}$) | Reference |
|---|---|---|---|---|
| PdFe$_3$O$_4$C | 70° C; 2 M ethanol in 1 M KOH | 0.5 | 31 | This work |
| PdSnO$_2$C | 70° C; 2 M ethanol in 1 M KOH | 0.5 | 28 | This work |
| PdFe$_3$O$_4$SnO$_2$/C | 70° C; 2 M ethanol in 1 M KOH | 0.5 | 30 | This work |
| PdNi/EGO | 50° C; 1 M ethanol in 1 M NaOH | 1.0 | 16.6 | [101] |
| Pd$_1$Nb$_1$/C | 60° C; 2 M ethanol in 1 M KOH | 1.0 | 27 | [102] |
| Pd$_{90}$Sn$_{10}$/C | 75 °C; 2 M ethanol in 2 M KOH | 1.0 | 15 | [96] |
| Pd$_3$Ru/C | 60° C; 3 M ethanol in 3 M KOH | 2.0 | 123 | [103] |
| Pd$_1$Sn$_3$/Vulcan XC-72 | 80° C; 2 M ethanol in 1 M KOH | 1.0 | 42 | [4] |

*EGO: (exfoliated graphene oxide)

## CONCLUSIONS

This study aimed to modify Pd-based electrocatalysts with Fe$_3$O$_4$ nano-octahedra and SnO$_2$ nanorods, and to evaluate their activity in the ethanol oxidation reaction (EOR) in alkaline medium. The synthesis by chemical reduction via sodium borohydride was a successful route to obtain palladium nanoparticles, as well as the hydrothermal routes used to obtain oxides with controlled morphology, which were supported on carbon by the impregnation method. The success of the synthetic routes was confirmed by physicochemical characterizations such as XRD, SEM-EDS, TEM, ICP-MS, and Raman spectroscopy. The SEM/EDS elemental mapping confirmed the presence of Pd, Fe, O and C in the respective electrocatalysts. The controlled morphology was confirmed with HR-TEM images, and the elemental composition was shown to be close to the initial nominal mass ratio using ICP-MS and EDS results. Raman spectroscopy measurements revealed an increased $I_D/I_G$ ratio in the modified nanostructured materials compared to Vulcan XC-72 carbon, suggesting a greater number of defects on the carbon surface. The XPS results showed that the $3d_{5/2}$ and $3d_{3/2}$ peaks of Pd for $PdFe_3O_4/C$, $PdSnO_2/C$, and $PdFe_3O_4SnO_2/C$ were shifted by ~0.5 eV to higher binding energies compared to Pd/C, indicating a loss of electron density in Pd due to strong metal–oxide interactions, which may suggest a better tolerance to catalytic poisoning. The binary and ternary materials also showed higher % of Pd oxides than the Pd/C materials, which indicates a greater supply of oxygenated species required for the oxidation of the EOR intermediates. The EOR studies showed that the $PdFe_3O_4/C$ material presented the highest mass activity (1426 mA mg$^{-1}$ Pd), approximately twice that obtained with the commercial Pd/C AA and simple Pd/C, followed by $PdSnO_2/C$ (1135 mA mg$^{-1}$ Pd) and $PdFe_3O_4SnO_2/C$ (1074 mA mg$^{-1}$ Pd). Stability studies using on-line SFC-ICP-M indicated that Fe may have contributed to the stability of the electrocatalyst since Pd dissolution was observed in the $PdFe_3O_4/C$ material and no Fe dissolution, corroborating the chronoamperometry studies. The enhanced activity and improved stability can be attributed to the bifunctional mechanism enabled by the presence of $Fe_3O_4$ nano-octahedra and $SnO_2$ nanorods, which facilitate the oxidation of reaction intermediates, corroborating the XPS results, reducing catalytic poisoning, and improving the activity for EOR. Notably, the $PdFe_3O_4/C$ anode electrocatalyst achieved the highest power density among the oxide-modified materials (31 mW cm$^{-2}$ at 70 °C), despite a ~45% reduction in palladium content compared to the commercial catalyst. Furthermore, an electronic effect may also be operating in $Fe_3O_4$-containing materials due to their magnetic properties, which may weaken the interaction strength of toxic intermediates such as CO.

**Acknowledgments**

The authors acknowledge financial support from the following Brazilian research financing institutions: Fundação de Amparo à Pesquisa do Estado de São Paulo (FAPESP, 2018/18675-8, 2021/10033-0, 2017/10118-0, 2017/21846-6, 2017/22976-0, 2017/26288-1, 2020/14100-0, 2022/15252-4, 2021/05364-7, 2021/14394-7, 2022/12895-1) and